\documentclass{article}
\usepackage[a4paper, total={6in, 8in}]{geometry} 
\usepackage{authblk} 
\usepackage{graphicx} 
\usepackage{amsthm} 
\newtheorem{Prop}{Proposition}
\newtheorem{Thm}{Theorem}
\newtheorem{Cor}{Corollary}
\newcommand{\ket}[1]{\left|#1\right>}

\title{Resource-compact time-optimal quantum computation}
\author{Taewan~Kim, Kyunghyun~Baek, Yongsoo~Hwang, and Jeongho~Bang}
\affil{\small Electronics and Telecommunications Research Institute, Daejeon, 34129, Korea\\ 
e-mail: TaewanKim@etri.re.kr}
\date{}

\begin{document}
\maketitle
\begin{abstract}
    Fault-tolerant quantum computation enables reliable quantum computation but incurs a significant overhead from both time and resource perspectives.
    To reduce computation time, Austin G. Fowler proposed time-optimal quantum computation by constructing a quantum circuit for a fault-tolerant $T$ gate without probabilistic $S$ gate correction.
    In this work, we introduce a resource-compact quantum circuit that significantly reduces resource requirements by more than 60\% for a fault-tolerant $T$ gate without probabilistic $S$ gate correction.
    Consequently, we present a quantum circuit that minimizes resource utilization for time-optimal quantum computation, demonstrating efficient time-optimal quantum computation.
    Additionally, we describe an efficient form involving initialization, CNOTs, and measurements, laying the foundation for the development of an efficient compiler for fault-tolerant quantum computation.
\end{abstract}
 
\section{Introduction}
Fault-tolerant quantum computing~\cite{Shor,AB,Gottesman} is necessary for reliable computation in real-world environments.
Overcoming errors and ensuring the reliability of quantum computations are critical steps towards achieving the transformative capabilities promised by quantum computing technology. 
Fault-tolerant quantum computation imposes significant resource demands, encompassing the need for additional qubits, increased circuit depth, ancillary qubits, and computational overhead for error correction. 
Addressing these challenges is essential for advancing the field of quantum computing system software and realizing the full potential of fault-tolerant quantum algorithms in practical applications.
Thus, it is necessary to reduce the resources required to run fault-tolerant quantum computation as much as possible.
For this purpose, several studies have been conducted and are still in progress~\cite{SPMH,AMMR,RS,KC,GMM}.

In fault-tolerant quantum computation, error correction and feedforward processes were believed to cause considerable time overhead, highlighting a key limitation in efficiency.
However, it is known that such overhead can be significantly reduced by a circuit that can perform $T$ gates without applying $S$ gate correction probabilistically~\cite{Fowler}. 
Thus, given a quantum error-correcting code that enables universal fault-tolerant quantum computation and transversal measurement of logical $X$ and $Z$, the methodology for executing {\it time-optimal quantum computation} has been proposed.
This means that the asymptotic time complexity of the arbitrary quantum algorithm execution corresponds to the product of the number of layers consisting of independent $T$ gates and a single physical measurement time.
In other words, surprisingly, it was demonstrated that the number of independent $T$ gates can determine the overall execution speed of quantum computation, highlighting its importance~\cite{GMM,BMNG+,CC,HEMN,AMM}.

On the other hand, arbitrary fault-tolerant quantum computation can be converted into the Initialization, CNOT, and Measurement form~\cite{PPND}. 
The form is called {\it ICM form}.
It is known that the form allows a more flexible approach towards circuit optimization for an appropriate compiler to produce a fault-tolerant, error-corrected description from a higher-level quantum circuit. 
The ICM form is composed based on the previous known circuit for fault-tolerant implementation for the $T$ gate~\cite{Fowler}. 

The previous known circuit for fault-tolerant $T$ gate implementation is based on selective destination teleportation and selective source teleportation. 
Thus, the circuit requires more resources than are absolutely necessary to perform it.
In this work, we design a circuit that directly performs fault-tolerantly $T$ gate without selective destination teleportation and selective source teleportation in order to design a more efficient circuit.
Utilizing our circuit design for a fault-tolerant $T$ gate, we demonstrate time-optimal quantum computation with minimized resources and describe the ICM form, also optimized in terms of resources.
 
\section{Quantum circuits for a fault-tolerant $T$ gate}
An arbitrary fault-tolerant quantum computation can be performed using only controlled-$NOT$, $H$, $S$, and $T$ gates~\cite{NC}. 
Generally, it has been turned out to be very simple to implement the controlled-$NOT$, $H$, and $S$ gates fault-tolerantly.
In order to complete the set of gates for universal quantum computation, it is necessary to perform the $T$ gate in a fault-tolerant manner.

\begin{figure}
    \centering
    \includegraphics[width=0.35\textwidth]{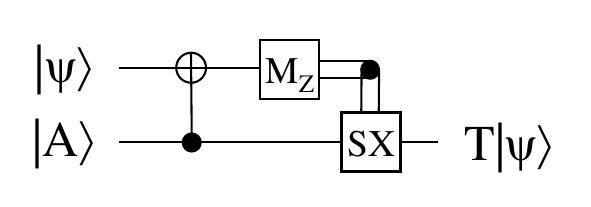} 
    \caption{Quantum circuit for fault-tolerant $T$ gate with probabilistic $S$ gate correction~\cite{NC}.}
    \label{fig:FT_T_with_S}
\end{figure}

A quantum circuit implementing fault-tolerantly a $T$ gate is shown in Figure~\ref{fig:FT_T_with_S}.
Here, $\ket{A}=TH\ket{0}$.
The state $\ket{A}$ can be prepared either directly via state distillation~\cite{BK}.
In the circuit, the measurement is performed in the first qubit, and if the measurement result is 0 then it is done. 
Otherwise, the operation $SX$ is performed to the second qubit. 
In other words, based on the measurement results, the $S$ gate is applied probabilistically. 

\begin{figure}
    \includegraphics[width=1.00\textwidth]{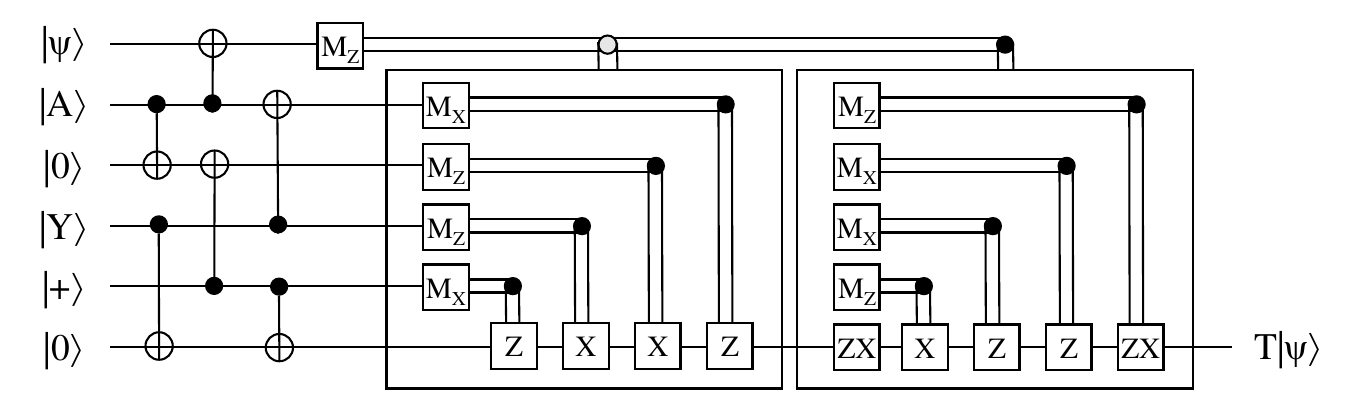} 
    \caption{Quantum circuit for fault-tolerant $T$ gate without probabilistic $S$ gate correction~\cite{Fowler}.}
    \label{fig:FT_T_without_S}
\end{figure}
A quantum circuit for a fault-tolerant $T$ gate without a probabilistic $S$ gate correction can be performed by the combination of selective destination teleportation, selective source teleportation and the circuit for a fault-tolerant $T$ gate in Figure~\ref{fig:FT_T_with_S}.
Figure~\ref{fig:FT_T_without_S} shows a quantum circuit implementing a $T$ gate without a probabilistic $S$ gate correction~\cite{Fowler}. 
Here, $\ket{Y}=SH\ket{0}$ and $\ket{+}=H\ket{0}$. 
The circuit shown in Figure~\ref{fig:FT_T_without_S} is slightly modified using circuit identity to reduce depth. 
To ensure that the $T$ gate is performed accurately, we have calculated operations based on the measurement results and included them in the Figure~\ref{fig:FT_T_without_S}. 
From the circuit depicted in the Figure~\ref{fig:FT_T_without_S}, we can deduce the following Proposition~\ref{prop:T}.

\begin{Prop}
    A quantum circuit to perform a $T$ gate without probabilistically applying the $S$ gate can be designed using a maximum of 5 ancillary qubits, 6 CNOTs and 5 $X$- or $Z$-basis measurements~\cite{Fowler}.
    \label{prop:T}
\end{Prop}
Since $\ket{Y}=SH\ket{0}$, the $S$ gate is included in the process of generating the $\ket{Y}$ state and is applied deterministically, not probabilistically.
In the circuit in Figure~\ref{fig:FT_T_without_S}, operations based on measurement results are performed using only the $X$ or $Z$ gate. 
Thus, performing $X$ or $Z$ gates followed by Clifford gates can be converted to performing Clifford gates first and then performing $X$ or $Z$ gates appropriately. 
It can be used to parallelize quantum circuits.

\begin{figure}
    \centering
    \includegraphics[width=0.6\textwidth]{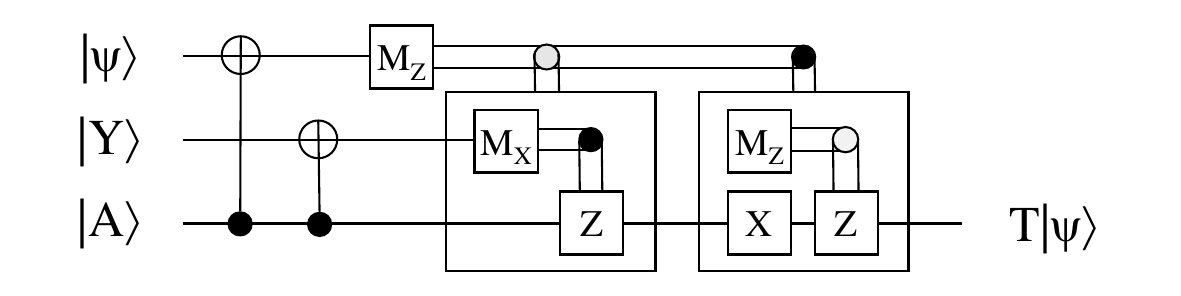} 
    \caption{Our quantum circuit for fault-tolerant $T$ gate without probabilistic $S$ gate correction.}
    \label{fig:Effi_FT_T_without_S}
\end{figure}
Constructing a quantum circuit for a fault-tolerant $T$ gate without employing probabilistic $S$ gate correction via selective destination and source teleportation may seem intuitively straightforward. 
However, we design a circuit that directly performs fault-tolerantly $T$ gate without selective destination and selective source teleportation in order to design a more efficient circuit.
Figure~\ref{fig:Effi_FT_T_without_S} shows our circuit. 
The circuit utilizes only two ancillary states, represented as $\ket{Y}$ and $\ket{A}$.
The circuit necessitates 2 CNOT operations and also requires 2 measurements in either the $X$- or $Z$-basis.
Therefore, from the circuit depicted in the Figure~\ref{fig:Effi_FT_T_without_S}, we can deduce the following Theorem~\ref{thm:T}.
\begin{Thm}
    A quantum circuit to perform a $T$ gate without probabilistically applying the $S$ gate can be designed using a maximum of 2 ancillary qubits, 2 CNOTs and 2 $X$- or $Z$-basis measurements.
    \label{thm:T}
\end{Thm}
The proof can be established through straightforward calculations. However, we elaborate on the proof to demonstrate the correctness of our circuit in implementing the $T$ gate.
\begin{proof}
Let $\ket{\psi} = \alpha\ket{0} + \beta\ket{1}$, where ${|\alpha|}^2 + {|\beta|^2} = 1$. Then,
\begin{eqnarray}
    \ket{\psi}\ket{Y}\ket{A} &=& \frac{1}{2}(\alpha\ket{000} + e^{\frac{\pi i}{4}}\alpha\ket{001} + i\alpha\ket{010} + ie^{\frac{\pi i}{4}}\alpha\ket{011} \nonumber\\
    & & + \beta\ket{100} + e^{\frac{\pi i}{4}}\beta\ket{101} + i\beta\ket{110} + ie^{\frac{\pi i}{4}}\beta\ket{111}).
\end{eqnarray}

Thus,
\begin{eqnarray}
    CNOT_{32}CNOT_{31}\ket{\psi}\ket{Y}\ket{A} &=& \frac{1}{2}(\alpha\ket{000} + e^{\frac{\pi i}{4}}\alpha\ket{111} + i\alpha\ket{010} + ie^{\frac{\pi i}{4}}\alpha\ket{101} \nonumber\\
    & & + \beta\ket{100} + e^{\frac{\pi i}{4}}\beta\ket{011} + i\beta\ket{110} + ie^{\frac{\pi i}{4}}\beta\ket{001}).
\end{eqnarray}

Next, a $Z$-basis measurement is performed on the first qubit, and depending on the measurement result of the first qubit, $X$-basis measurement or $Z$-basis measurement is performed on the second qubit.
Let us denote the outcome of the $Z$-basis measurement result on the first qubit as $m_{Z_{1}}$. Let us denote the outcome of $X(Z)$-basis measurement result on the second qubit as $m_{X_{2}}(m_{Z_{2}})$.

\begin{enumerate}
    \item If $m_{Z_{1}} = 0$, then $\frac{1}{\sqrt{2}}\ket{0}(\alpha\ket{00} + i\alpha\ket{10} + e^{\frac{\pi i}{4}}\beta\ket{11} + ie^{\frac{\pi i}{4}}\beta\ket{01})$.
        \begin{enumerate}
            \item If $m_{X_{2}} = 0$, then $\frac{1+i}{\sqrt{2}}\ket{0}\ket{+}T_{3}\ket{\psi}$.
            \item If $m_{X_{2}} = 1$, then $\frac{1-i}{\sqrt{2}}\ket{0}\ket{-}(\alpha\ket{0} - e^{\frac{\pi i}{4}}\beta\ket{1})$.
        \end{enumerate}
        Therefore, $\frac{1-i}{\sqrt{2}}\ket{0}\ket{-}Z_{3}(\alpha\ket{0} - e^{\frac{\pi i}{4}}\beta\ket{1}) = \frac{1-i}{\sqrt{2}}\ket{0}\ket{-}T_{3}\ket{\psi}$.
    \item If $m_{Z_{1}} = 1$, then $\frac{1}{\sqrt{2}}\ket{1}(e^{\frac{\pi i}{4}}\alpha\ket{11} + ie^{\frac{\pi i}{4}}\alpha\ket{01} + \beta\ket{00} + i\beta\ket{10})$.
        \begin{enumerate}
            \item If $m_{Z_{2}} = 0$, then $\ket{10}(ie^{\frac{\pi i}{4}}\alpha\ket{1} + \beta\ket{0})$.\\
            Therefore, $\ket{10}Z_{3}X_{3}(ie^{\frac{\pi i}{4}}\alpha\ket{1} + \beta\ket{0}) = ie^{\frac{\pi i}{4}}\ket{10}T_{3}\ket{\psi}$.
            \item If $m_{Z_{2}} = 1$, then $\ket{11}(e^{\frac{\pi i}{4}}\alpha\ket{1} + i\beta\ket{0})$.\\
            Therefore, $\ket{11}X_{3}(e^{\frac{\pi i}{4}}\alpha\ket{1} + i\beta\ket{0}) = e^{\frac{\pi i}{4}}\ket{11}T_{3}\ket{\psi}$.
        \end{enumerate}
\end{enumerate}
\end{proof}

\begin{table}
    \centering
    \begin{tabular}{c c c}
    \hline\hline
    resources & the previous circuit & our circuit\\
    \hline
    number of ancillary qubits & 5 & 2\\
    number of $CNOT$s & 6 & 2\\
    number of measurements & 5 & 2\\
    \hline\hline
    \end{tabular}
    \caption{Comparison between the previous circuit and our circuit for fault-tolerant $T$ gate implementation.
    Reduction rates: Ancillary qubits - $60\%$, CNOTs - $67\%$, Measurements - $60\%$.}
    \label{tab:Comparison}
\end{table}
Table~\ref{tab:Comparison} presents a comparison between the previously known circuit and our proposed circuit for implementing the fault-tolerant $T$ gate.
For fault-tolerant quantum computation employing $n$ independent $T$ gates, the previous circuit necessitates $5n$ ancillary qubits, whereas our circuit requires only $2n$. 
Consequently, relative to the prior approach, the ratio of ancillary qubits required in the circuit can be reduced by $60\%$. 
The number of CNOTs and number of measurements for fault-tolerant quantum computation can also be reduced significantly.
Furthermore, operations based on measurement results are considerably simpler than those of previous circuits.

A circuit for a fault-tolerant $T$ gate appears that a minimum of two ancillary qubits are essential for deterministic fault-tolerant implementation of the $T$ gate without probabilistic $S$ gate correction. 
One corresponds to the state $\ket{A}$ for the application of the $T$ gate, while the other relates to the state $\ket{Y}$ intended for the $S$ gate application. 
Each of these ancillary qubits seemingly necessitates at least one two-qubit gate to entangle with a given state $\ket{\psi}$.
Thus, configuring a quantum circuit for fault-tolerant $T$ gate implementation appears challenging, particularly when aiming for further resource reduction compared to our circuit, especially under the same conditions where measurements and operations depending on the measurement results.

\begin{figure}
    \centering
    \includegraphics[width=0.6\textwidth]{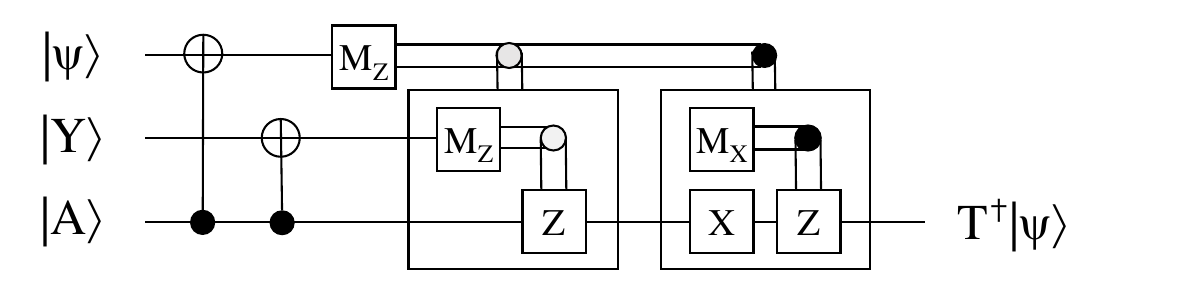} 
    \caption{Our quantum circuit for fault-tolerant $T^{\dagger}$ gate without probabilistic $S$ gate correction.}
    \label{fig:Effi_FT_Tdag_without_S}
\end{figure}
In a similar way, a quantum circuit for fault-tolerant $T^{\dagger}$ gate implementation can be obtained as shown in Figure~\ref{fig:Effi_FT_Tdag_without_S}.
The structure of a quantum circuit for fault-tolerant $T^{\dagger}$ gate implementation is similar to a structure of a quantum circuit for fault-tolerant $T$ gate implementation. 

\section{Efficient time-optimal quantum computation}
Our proposed circuit for implementing the fault-tolerant $T$ gate can be used for all fault-tolerant quantum computation. 
The circuit $(HT)^{n}$ can be regarded as a simple example of a circuit previously thought to require substantial time overhead. However, the overhead can be greatly reduced by using the technique of time-optimal quantum computation~\cite{Fowler}.

\begin{figure}[ht]
    \centering
    \includegraphics[width=1.00\textwidth]{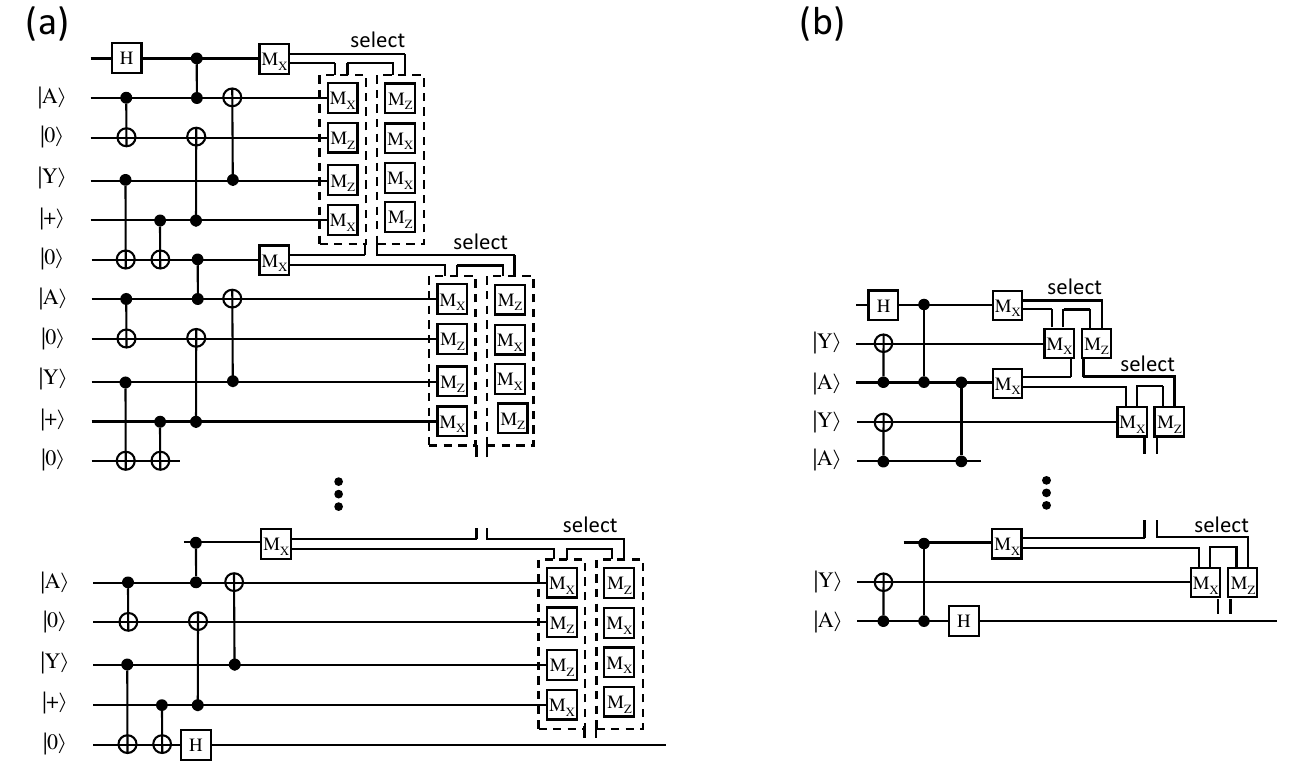} 
    \caption{Quantum circuits for fault-tolerant $(HT)^{n}$:
    (a) Using the previous known circuit
    (b) Using our circuit.
    Reduction rates: Ancillary qubits - $60\%$, CNOTs - $80\%$, Measurements - $60\%$.}
    \label{fig:FT_HTn}
\end{figure}
The circuit $(HT)^{n}$ is shown in Figure~\ref{fig:FT_HTn} (a). 
The circuits in Figure~\ref{fig:FT_HTn} was expressed according to Fowler's notation~\cite{Fowler}. 
The circuit in Figure~\ref{fig:FT_HTn} (a) was slightly modified using circuit identity to reduce depth. 
It consists of $5n$ ancillary qubits, $5n$ CNOTs, $n$ CZ, 2 $H$s, and $5n$ $X$- or $Z$-basis measurements.
The circuit depth encompasses a depth of 4 for the Clifford circuit and an additional depth of $n+1$ for measurements, resulting in a total depth of $n+4$ excluding $X$ or $Z$ gates based on measurement results.
In principle, only the time-ordered $X$- or $Z$-basis measurements constrain the speed of quantum computation, as all gates preceding measurements can be executed in constant time.

Utilizing our proposed design, the circuit $(HT)^{n}$ can be more efficiently constructed, as depicted in Figure~\ref{fig:FT_HTn} (b).
It consists of $2n$ ancillary qubits, $n$ CNOTs, $n$ CZ, 2 $H$s and $2n$ $X$- or $Z$-basis measurements.
The circuit depth encompasses a depth of 3 for the Clifford circuit and an additional depth of $n+1$ for measurements, resulting in a total depth of $n+4$ excluding $X$ or $Z$ gates based on measurement results.
The ratio of ancillary qubits needed for the fault-tolerant $(HT)^{n}$ circuit has decreased by $60\%$, from $5n$ to $2n$.
Although the ratio of CZs required is the same at $n$, the ratio of CNOTs has decreased by $80\%$ from $5n$ to $n$.
The ratio of measurements required has also decreased by $60\%$, from $5n$ to $2n$.

\begin{figure}[ht]
    \includegraphics[width=1.00\textwidth]{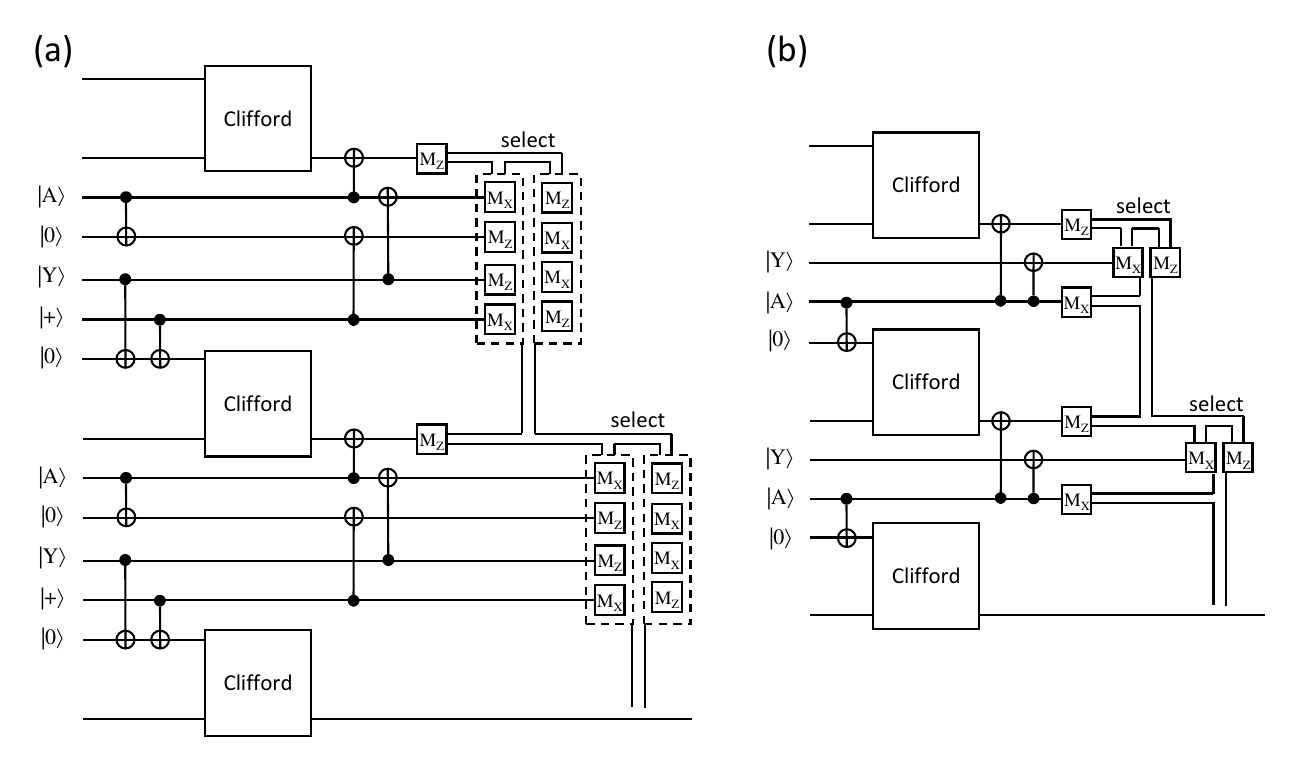}
    \caption{Time-optimal quantum computation:
    (a) Using the previous known circuit
    (b) Using our circuit.
    Reduction rates: Ancillary qubits - $40\%$, CNOTs - $50\%$, Measurements - $40\%$.}
    \label{fig:TOQC}
\end{figure}
In general, arbitrary quantum computations can be consisted of Clifford gates and $T$ gates. 
The circuit in the Figure~\ref{fig:TOQC} (a) can be obtained by using the previous circuit for fault-tolerant $T$ gate implementation.
The circuit consists of parallel Clifford gates including CNOTs for $T$ gate implementation and time-ordered $X$- or $Z$-basis measurements.
This is known as {\it time-optimal quantum computation}.

Utilizing our proposed design, the circuit can be more efficiently constructed, as depicted in Figure~\ref{fig:TOQC} (b). 
For each fault-tolerant $T$ gate implementation,
the number of ancillary qubits required is only 2.
However, to parallel Clifford gates, gate teleportation is additionally performed.
Thus, one additional ancillary qubit is used for each $T$ gates.
Even considering parallel Clifford gates, the ratio of ancillary qubits required for time-optimal quantum computation is reduced by $40\%$.
If parallel Clifford gates are not considered, quantum computation is possible with fewer resources because gate teleportation is not required.
The following Corollary~\ref{cor:T_depth} can be obtained.
\begin{Cor}
    Arbitrary quantum computation can be transformed into a quantum circuit for time-optimal quantum computation using a maximum of 3 ancillary qubits, 3 CNOT operations, 3 measurements in either the $X$- or $Z$-basis, and solely $X$ or $Z$ operations depending on the measurement results for each independent $T$ gate.
    \label{cor:T_depth}
\end{Cor}
As evident from the Corollary~\ref{cor:T_depth}, executing fault-tolerant quantum computation expeditiously is feasible with fewer resources compared to existing methods.

\section{Efficient ICM form}
Arbitrary quantum computation can be converted into the Initialization, CNOT, and Measurement form~\cite{PPND}. First, the initialization layer of qubits consists of one of four distinct states ($\ket{0}$, $\ket{+}$, $\ket{Y}$, $\ket{A}$). 
Secondly, the CNOT layer consists of a massive and deterministic array of CNOT operations. 
Last, the measurement layer consists of a series of time-ordered $X$- or $Z$-basis measurements.
The ICM form facilitates a versatile methodology for circuit optimization. 
Concurrently, the package yields either a standard circuit or a canonical geometric description, essential for interfacing with contemporary hardware architectures that employ topological quantum codes.

\begin{figure}
    \centering
    \includegraphics[width=0.6\textwidth]{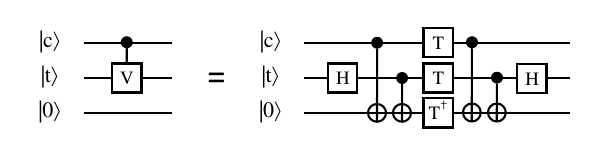}
    \caption{Our decomposition of the controlled-$V$ gate employing an ancillary qubit.}
    \label{fig:decomp_CV}
\end{figure} 
As a simple example, a controlled-$V$ gate can be converted into the ICM form.
First of all, the controlled-$V$ gate can be decomposed into 2 $H$s, 4 CNOTs, 2 $T$s and 1 $T^{\dagger}$ with an ancillary qubit as shown in Figure~\ref{fig:decomp_CV}. 
\begin{figure}
    \centering
    \includegraphics[width=0.5\textwidth]{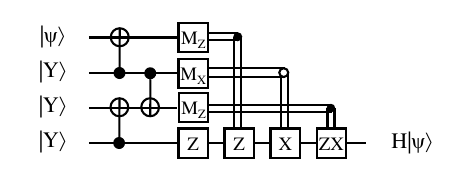}
    \caption{ICM form for the $H$ gate~\cite{PPND2}.}
    \label{fig:FT_H}
\end{figure} 
Through simple calculations, the $H$ gate can be converted into the ICM form as shown in Figure~\ref{fig:FT_H}~\cite{PPND2}.
\begin{figure}[t]
    \includegraphics[width=1.00\textwidth]{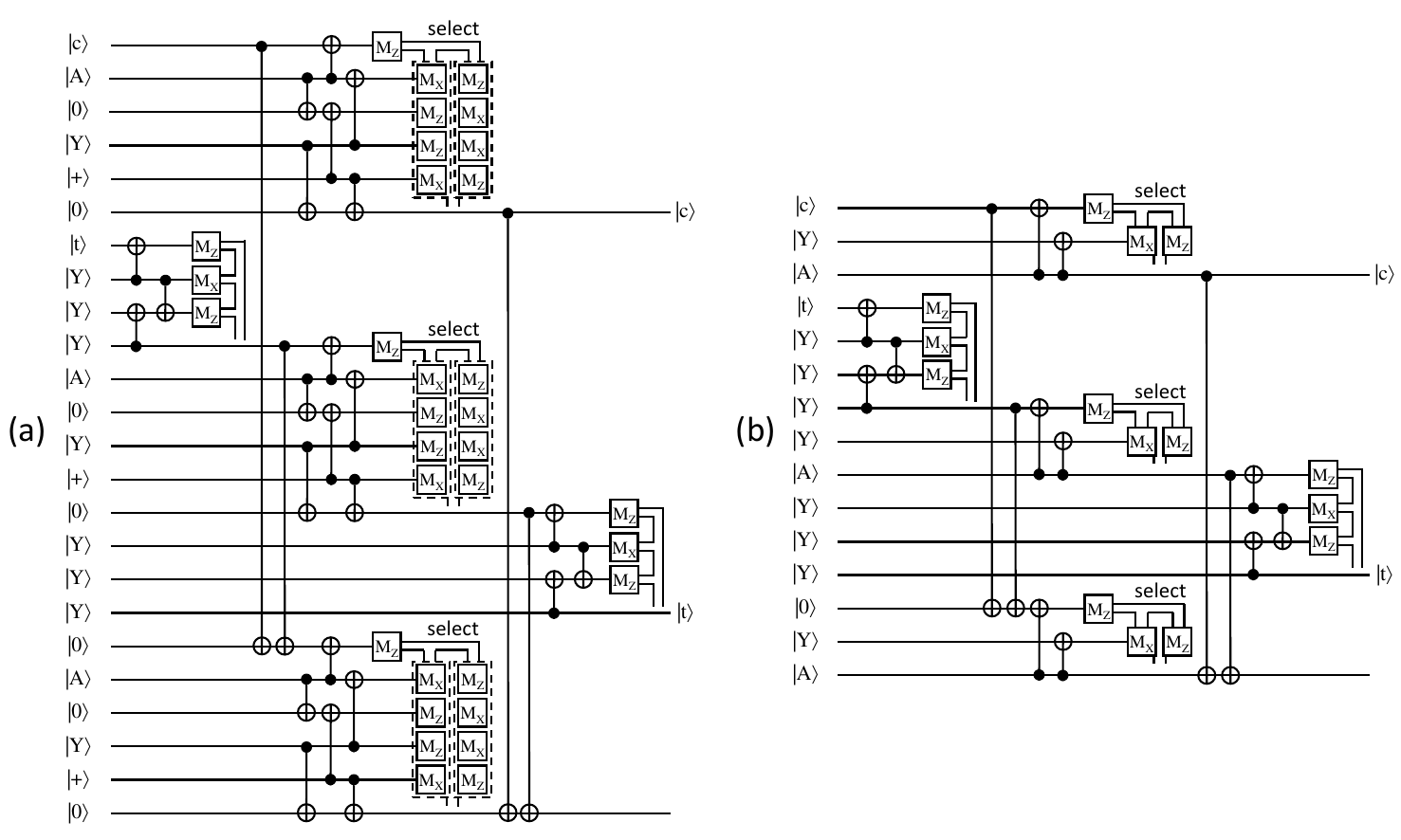}
    \caption{ICM form of the controlled-$V$ gate:
    (a) ICM form using the previous known circuit for the controlled-$V$ gate.
    (b) Our ICM form utilizing our circuit for the controlled-$V$ gate. Reduction rates: Ancillary qubits - $41\%$, CNOTs - $43\%$, Measurements - $43\%$.}
    \label{fig:CV}
\end{figure}
By using the previous circuit for the implementation of the $T$ gate, circuit shown in Figure~\ref{fig:CV} (a) can be obtained.
It consists of 22 ancillary qubits, 28 CNOTs and 21 $X$- or $Z$-basis measurements.
By employing our circuit for $T$ gates depicted in Figure~\ref{fig:Effi_FT_T_without_S} and the circuit for $T^{\dagger}$ gates shown in Figure~\ref{fig:Effi_FT_Tdag_without_S}, the form can be designed with reduced resources. The ICM form for the controlled-$V$ gate is illustrated in Figure~\ref{fig:CV} (b).
It consists of 13 ancillary qubits, 16 CNOTs and 12 $X$- or $Z$-basis measurements.
Therefore, it can be seen that all resources are reduced by more than 40$\%$ compared to using a circuit in Figure~\ref{fig:CV} (a).
Consequently, it can facilitate the development of an efficient compiler for fault-tolerant quantum computation.

\section{Conclusion}
We have explored the expeditious execution of fault-tolerant quantum computation while minimizing resource consumption.
We have presented a quantum circuit that significantly reduces resource requirements by more than 60\% for a fault-tolerant $T$ gate without probabilistic $S$ gate correction.
Our circuit exhibits a 60\% reduction in ancillary qubits, a 67\% reduction in CNOT gates, and a 60\% reduction in measurements.
Consequently, we have presented a quantum circuit that minimizes resource utilization for time-optimal quantum computation, demonstrating efficient time-optimal quantum computation with over 40\% reduced resources.
Specifically, ancillary qubits are reduced by 40\%, CNOT gates by 50\%, and measurements by 40\%. 
Additionally, we have described the efficient ICM form for development of a compiler for fault-tolerant quantum computation.

The efficiency of our circuit has been attained by directly configuring the circuit without relying on selective destination teleportation and selective source teleportation while attempting to implement a fault-tolerant $T$ gate without probabilistically applying the $S$ gate.
Configuring a quantum circuit for fault-tolerant $T$ gate implementation appears challenging, especially when aiming for further resource reduction compared to our circuit under the same measurement conditions.

Our study emphasizes a significant enhancement in the efficiency of fault-tolerant $T$ gate implementation. 
The circuit holds potential for contributing to the advancement of quantum computing system software, such as the development of an efficient compiler for fault-tolerant quantum computation.

\section*{Acknowledgements}
T.~Kim expresses gratitude to Professor Soojoon Lee and Professor Hayata Yamasaki for their valuable discussions and insightful comments.
This work was supported by the Ministry of Science, ICT and Future Planning (MSIP) by the Institute of Information and Communications Technology Planning and Evaluation grant funded by the Korean government (2019-0-00003, ``Research and Development of Core Technologies for Programming, Running, Implementing and Validating of Fault-Tolerant Quantum Computing System") and the National Research Foundation of Korea (NRF-2021M3E4A1038213, RS-2023-00281456). 



\begin{thebibliography}{1}
    \bibitem{Shor} P.~W.~Shor, {\em Fault-tolerant quantum computation}, In Proceedings of 37th Conference on Foundations of Computer Science, pp. 56-65. IEEE, 1996.
    \bibitem{AB} D.~Aharonov and M. Ben-Or, {\em Fault-tolerant quantum computation with constant error}, In Proceedings of the Twenty-Ninth Annual ACM Symposium on Theory of Computing, pp. 176-188, 1997.
    \bibitem{Gottesman} D.~Gottesman, {\em An introduction to quantum error correction and fault-tolerant quantum computation}, in Quantum Information Science and Its Contributions to Mathematics, Proceedings of Symposia in Applied Mathematics, {\bf 68} pp. 13-58 (2010). 
    \bibitem{SPMH} V.~V.~Shende, A.~K.~Prasad, I.~L.~Markov and J.~P.~Hayes, {\em Synthesis of reversible logic circuits}, IEEE Transactions on Computer-Aided Design of Integrated Circuits and Systems {\bf 22} pp.710-722, 2003.
    \bibitem{AMMR} M.~Amy, D.~Maslov, M.~Mosca and M.~Roetteler, {\em A meet-in-the-middle algorithm for fast synthesis of depth-optimal quantum circuits}, IEEE Transactions on Computer-Aided Design of Integrated Circuits and Systems {\bf 32} pp. 818-830, 2013.
    \bibitem{RS} N.~J.~Ross and P.~Selinger, {\em Optimal ancilla-free Clifford+T approximation of z-rotations}, Quantum Information and Computation {\bf 16}, 901 (2016).
    \bibitem{KC} T.~Kim and B.-S.~Choi, {\em Efficient decomposition methods for controlled-$R_{n}$ using a single ancillary qubit}, Scientific Reports {\bf 8}, 5445 (2018).
    \bibitem{GMM} V.~Gheorghiu, M.~Mosca and P.~Mukhopadhyay, {\em T-count and T-depth of any multi-qubit unitary}, npj Quantum Information {\bf 8}, 141 (2022).
    \bibitem{Fowler} A.~G.~Fowler, {\em Time-optimal quantum computation}, arXiv:1210.4626 (2012).
    \bibitem{BMNG+} R.~Babbush, J.~R.~McClean, M.~Newman, C.~Gidney, S.~Boixo, and H.~Neven, {\em Focus beyond quadratic speedups for error-corrected quantum advantage}, PRX Quantum {\bf 2}, 010103 (2021).
    \bibitem{CC} C.~Chamberland and E.~T.~Campbell, {\em Universal quantum computing with twist-free and temporally encoded lattice surgery}, PRX Quantum {\bf 3}, 010331 (2022).
    \bibitem{HEMN} M.~Hanks, M.~P.~Estarellas, W.~J.~Munro, and K.~Nemoto, {\em Effective compression of quantum braided circuits aided by ZX-calculus}, Phys. Rev. X {\bf 10}, 041030 (2020).
    \bibitem{AMM} M.~Amy, D.~Maslov, and M.~Mosca, {\em Polynomial-time T-Depth optimization of Clifford+T circuits via matroid partitioning}, IEEE Transactions on Computer-Aided Design of Integrated Circuits and Systems {\bf 33} pp.1476-1489 (2014)
    \bibitem{PPND} A.~Paler, I.~Polian, K.~Nemoto and S.~J.~Devitt, {\em Fault-tolerant high level quantum circuits: form, compilation and description}, Quantum Science and Technology, {\bf 2}, 025003 (2017).
    \bibitem{NC} M.~Nielsen and I.~Chuang, {\em Quantum Computation and Quantum Information}, Cambridge University Press, 2000.
    \bibitem{BK} S.~Bravyi and A.~Kitaev, {\em Universal quantum computation with ideal Clifford gates and noisy ancillas}, Physical Review A {\bf 71}, 022316 (2005).
    \bibitem{PPND2} A.~Paler, I.~Polian, K.~Nemoto and S.~J.~Devitt, {\em A regular representation of quantum circuits}, Reversible Computation, Lecture Notes in Computer Science (LNCS) Krivine, Jean and Stefani, Jean-Bernard {\bf 9138} pp.139-154 (2015).
\end{thebibliography}
\end{document}